\documentclass[12pt]{article}
\usepackage{amsfonts}
\usepackage{amssymb}
\usepackage{amsbsy}
\usepackage{graphics}
\usepackage{epsfig}

\input epsf.sty

\newlength{\TZ}
\newcommand{\BEQ}{\begin{equation}}     
\newcommand{\BEA}{\begin{eqnarray}}
\newcommand{\EEQ}{\end{equation}}       
\newcommand{\EEA}{\end{eqnarray}}
\newcommand{\eps}{\varepsilon}          
\newcommand{\D}{{\rm d}}                
\newcommand{\II}{{\rm i}}               
\newcommand{\demi}{\frac{1}{2}}         
\newcommand{\wht}[1]{\widehat{#1}}      

\newcommand{\appsektion}[1]{\setcounter{equation}{0}\setcounter{subsection}{0}
\section*{Appendix. #1}
\renewcommand{\theequation}{A.\arabic{equation}}
              \renewcommand{\thesection}{A} }

\catcode`\@=11
\def\numberbysection{\@addtoreset{equation}{section}
        \def\theequation{\thesection.\arabic{equation}}}
\numberbysection

\begin{document}

\begin{titlepage}

\vskip 1.5 cm
\begin{center}
{\LARGE \bf On non-local representations of the ageing algebra
}
\end{center}

\vskip 2.0 cm
\centerline{{\bf Malte Henkel}$^a$ and {\bf Stoimen Stoimenov}$^b$}
\vskip 0.5 cm
\centerline{$^a$ Groupe de Physique Statistique, D\'epartement de Physique de la Mati\`ere et des Mat\'eriaux,}
\centerline{Institut Jean Lamour,
Nancy Universit\'e (UMR 7198 -- CNRS -- UHP -- INPL -- UPVM)}
\centerline{B.P. 70239, F -- 54506 Vand{\oe}uvre l\`es Nancy Cedex, France}
\vspace{0.5cm}
\centerline{$^b$ Institute of Nuclear Research and Nuclear Energy, Bulgarian Academy of Sciences,}
\centerline{72 Tsarigradsko chaussee, Blvd., BG -- 1784 Sofia, Bulgaria}

\begin{abstract}
The ageing algebra is a local dynamical symmetry of many ageing systems, far from equilibrium, and with
a dynamical exponent $z=2$. Here, new representations for an integer dynamical exponent $z=n$ are constructed, which
act non-locally on the physical scaling operators. The new mathematical mechanism which makes the infinitesimal generators
of the ageing algebra dynamical symmetries, is explicitly discussed for a $n$-dependent family of linear equations of
motion for the order-parameter. Finite transformations are derived through the exponentiation of the
infinitesimal generators and it is proposed to interpret them in terms of the transformation of distributions
of spatio-temporal coordinates. The two-point functions which transform co-variantly under the
new representations are computed, which quite distinct forms for $n$ even and $n$ odd. Depending on the sign
of the dimensionful mass parameter, the two-point scaling functions either decay monotonously or in an oscillatory
way towards zero. 
\end{abstract}

\end{titlepage}

\setcounter{footnote}{0}

\section{Introduction}

Non-relativistic space-time transformations have recently met with a lot of interest -- in
addition to fields such as hydrodynamics \cite{Ovsiannikov82,Zhang10},
they have been playing an increasing r\^ole in the analysis of
the long-time behaviour of strongly interacting many-body systems far from equilibrium \cite{Cugliandolo02,Henkel10}
and even more recently have arisen in non-relativistic limits of the AdS/CFT correspondence, see e.g.
\cite{Bala08,Minic08,Bagchi09c,Jottar10}, with interesting applications to cold atoms \cite{Son08,Fuertes09}.
One of the ingredients for physically interesting sets of space-time transformations appears to be some kind
of conformal invariance and recently, a classification of non-relativistic conformal space-time transformations was
presented \cite{Duval09}. An important sub-set of these is characterised by a finite dynamical exponent $0<z<\infty$.
Indeed, the list of sets of admissible generators which close into a Lie algebra is a rather short one,
in $d+1$ space-time dimensions:
\begin{enumerate}
\item the {\em conformal algebra} itself, in $d+1$ dimensions, with $z=1$.
\item the {\em Schr\"odinger algebra}, which was discovered by Lie in 1881 as a dynamical symmetry algebra of the $1D$ free
diffusion equation.\footnote{Jacobi already wrote down {\it en passant} in 1842/43 the elements of the corresponding Lie
group as symmetries of the Hamilton-Jacobi equation of a particle with an inverse-square potential, see \cite{Havas78}.}
The dynamical exponent is $z=2$.
This algebra was rediscovered in physics as a symmetry of free non-relativistic particles several times around 1970,
including \cite{Kastrup68,Hagen72,Niederer72,Jackiw72}. Non-linear examples of Schr\"odinger-invariant equations include the
Navier-Stokes equation \cite{Ovsiannikov82,Hassaine00,ORaif01} or Burger's equation \cite{Niederer78,Ivash97}, see e.g.
\cite{Henkel10} and refs. therein for a short summary of further examples. Schr\"odinger symmetry (or rather its sub-algebra
with time-translations left out) also arises in
the far-from-equilibrium dynamics of statistical systems \cite{Henkel94}, for example in simple magnets quenched to a
temperature $T<T_c$ below the critical temperature $T_c>0$ from a fully disordered initial state when $z=2$ is known
\cite{Bray94a,Cugliandolo02,Henkel10}.
\item the {\em conformal galilean algebra} {\sc cga}$(d)$ appears to have been first identified in \cite{Havas78},
but was independently
rediscovered in different contexts \cite{Henkel97,Negro97}. It is usually obtained, by a contraction,
as the non-relativistic limit of the
$(d+2)$-dimensional conformal algebra (itself obtained by a non-relativistic holographic construction)
\cite{Henkel03a,Martelli09,Bagchi09a,Bagchi09b,Leigh09,Leigh10,Jottar10}. In $d=1$
space dimension, there exists an infinite-dimensional extension which can be constructed from a contraction of a
pair of commuting Virasoro algebras \cite{Henkel02,Henkel06b,Bagchi09b}. In most representations, one has $z=1$, but
representations with $z=2$ are also known \cite{Henkel06b}.
\item in $d=2$ space dimensions, there exists the {\em exotic conformal galilean algebra} {\sc ecga},
which is the central extension  of the non-semi-simple {\sc cga}$(2)$ \cite{Lukierski06}.
Although one may readily identify linear equations invariant under {\sc ecga} \cite{Martelli09},
the construction of invariant non-linear equations is not straightforward \cite{Zhang10,Cherniha09}.
The dynamical exponent is $z=1$.
See \cite{Horvathy10} for a recent review and the relationship with non-commutative mechanics.
\item finally, there exists for $d=1$ a closed algebra with $z=\frac{3}{2}$ \cite{Henkel02}.
It is not yet clear how this might
fit into the general scheme of \cite{Duval09}, since it does not contain the full conformal structure and furthermore
its generators contain {\em fractional} space derivatives.
\end{enumerate}
This short list illustrates the practical difficulty of constructing sets of `conformal' space-time transformations for
a generic dynamical exponent $z\ne 1,2$ such that a Lie algebra is obtained. It is at present not fully
understood how to construct a dynamical symmetry even for a simple linear equation of the form (where $z\ne 1,2$)
\BEQ \label{1}
{\cal S}\psi(t,r) := \bigl( z \mu \partial_t - \partial_r^z \bigr) \psi(t,r) = 0
\EEQ
which arises as one of the most simple equations of motion of the order-parameter in studies of ageing far from
equilibrium \cite{Cannas01,Baumann07b}. Indeed, current attempts to find further dynamical symmetries of eq.~(\ref{1})
beyond the obvious translation-, dilatation- and rotation-symmetries (if $d>1$) only succeed at the price that
the further generators must be required to vanish on certain states (which are then declared to be the `physical' ones)
\cite{Henkel97,Henkel02,Henkel10}. Furthermore, these generators cannot, in general, be expressed in terms of first-order
differential operators (`vector fields' in mathematical terminology). In the context of statistical physics, the order-
parameter does not really satisfy a deterministic equation, but rather the r.h.s of (\ref{1}) is replaced by a random noise
term, which leads to a Langevin equation. However,
since the non-relativistic algebras mentioned above are all non-semi-simple
and their representations are projective, it is possible to study first the symmetries of the deterministic equation
(\ref{1}) and then use the resulting Bargman super-selection rules \cite{Bargman54} in order to reduce the calculation
of any average to the calculation  of averages within the deterministic part of the theory as defined by (\ref{1})
\cite{Picone04}. This procedure works not only for thermal noises and a simple diffusion equation with $z=2$, but can be
generalised to generic values of $z$ and fairly general noises, such as they may arise in reaction-diffusion systems
\cite{Baumann05b,Roethlein06,Baumann07,Baumann07b,Durang09}, see \cite{Henkel10} for a systematic presentation.

In this paper, we shall explore properties of a new kind of representations of the common sub-algebra $\mathfrak{age}(d)$
of the Schr\"odinger and conformal galilean algebras. For the sake of notational simplicity,
we shall restrict from now on to
$d=1$ spatial dimensions. Then the standard representation, on sufficiently differentiable space-time functions
$f(t,r)$, of  the Lie algebra $\mathfrak{age}(1):=\bigl\langle X_{0,1}, Y_{\pm \demi}, M_0\bigr\rangle$ is given by
\BEA
X_0 &=& -t\partial_t - \demi r \partial_r - \frac{x}{2} 
\hspace{3.15truecm} \mbox{\rm dilatation} \nonumber \\
X_1 &=& - t^2\partial_t - t r \partial_r - \frac{\cal M}{2} r^2 - (x+\xi)t 
\hspace{0,5truecm}\mbox{\rm special transformation}\nonumber \\
Y_{-\demi} &=& - \partial_r 
\hspace{5.42truecm} \mbox{\rm space-translations} \label{2} \\
Y_{\demi} &=& - t\partial_r - {\cal M}r 
\hspace{4.02truecm} \mbox{\rm Galilei-transformation}\nonumber \\
M_0 &=& -{\cal M} \hspace{5.27truecm} \mbox{\rm phase shift} \nonumber
\EEA
See (\ref{2.2}) for the commutators. This representation 
is characterised by the `mass' $\cal M$ and the pair of scaling dimensions $(x,\xi)$ whose values depend on the
scaling operator on which these generators act. If one defines the Schr\"odinger operator
\BEQ \label{3}
{\cal S} := 2{\cal M} \partial_t - \partial_r^2 + 2{\cal M}\left( x+\xi-\demi\right) \frac{1}{t}
\EEQ
then the equation ${\cal S}\psi(t,r)=0$ has $\mathfrak{age}(1)$ as dynamical symmetry, because of the
commutation relations
\BEQ
\bigl[ {\cal S}, Y_{\pm \demi}\big] \:=\: \bigl[ {\cal S}, M_0 \bigr] \:=\: 0 \;\; , \;\;
\bigl[ {\cal S}, X_0 \bigr] \:=\:  -{\cal S} \;\; , \;\; \bigl[{\cal S}, X_1\bigr] = -2t {\cal S}
\EEQ
which imply that any solution of ${\cal S}\psi=0$ is mapped onto another solution of the same equation. A physical
example for (\ref{3}) is given by the relaxation kinetics of the spherical model, or equivalently the $N\to\infty$ limit
of the O($N$) model, after a quench to a temperature $T\leq T_c$ at or below its critical temperature $T_c>0$
\cite{Godreche00b}.
The representation (\ref{2}) has a dynamical exponent $z=2$ and acts {\em locally}
on the space-time coordinates. We point out
that because time-translations (with a generator $X_{-1}=-\partial_t$) are not included and hence a system with
an $\mathfrak{age}$-symmetry is not at a stationary state, the scaling dimension $\xi$ arises as a further universal
characteristics of the relaxation process. By the transformation $\phi(t,r)=t^{\xi_{\phi}}\Phi(t,r)$, the physical
$\mathfrak{age}$-quasi-primary field $\phi$ with scaling dimensions $(x_{\phi},\xi_{\phi})$ can be related to the
Schr\"odinger-quasi-primary field $\Phi$ with scaling dimensions $(x_{\Phi}=x_{\phi}+2\xi_{\phi},\xi_{\Phi}=0)$
\cite{Henkel06a} and in
the transformed eq.~(\ref{3}) $x+\xi$ is replaced by $x_{\Phi}$. If this option is chosen, one must give up the
identity of physical and quasi-primary scaling operators, familiar from conformal invariance, which holds for
stationary, equilibrium systems. The potential term in (\ref{3}) can of course be eliminated by a similar transformation
or else by imposing the constraint $x+\xi=\demi$.

When trying to extend (\ref{2}) to a representation of {\sc cga}$(1)\supset \mathfrak{age}(1)$, 
the extra generators are not necessarily
first-order differential operators \cite{Henkel03a}. 
We believe that this fact should be taken seriously and its consequences studied. 
For this reason, and in order to find further representations of $\mathfrak{age}(1)$
with different values of $z$, we shall give in section~2 {\em non-local} representations of $\mathfrak{age}(1)$,
which admit any integer value $z=n\in\mathbb{N}$, but which cannot be reduced to first-order differential operators.
As we shall see, closure of these representations requires to restrict the representation space by performing a quotient
with respect to the Schr\"odinger equation ${\cal S}\psi=0$.
In section~3, we address the question how to interpret geometrically such infinitesimal generators by explicitly
constructing the finite space-time transformations given by the exponentiated
generators. The examples studied here suggest that these finite transformations 
might be viewed as transformations of {\em distributions}
of space-time coordinates, instead of precise transformations of the coordinates. An appendix compares this with the
finite Galilei-transformations. Next, in section~4, we derive the
co-variant two-point functions which depend strongly on the parity of $n$. We conclude in section~5.

\section{Non-local representation of the ageing algebra $\mathfrak{age}(1)$}
Consider a dynamical exponent with integer values $2\leq z=n\in\mathbb{N}$.
The generators of $\mathfrak{age}(1)$ we are interested in take the form
\BEA
X_0 &=& -\frac{n}{2}t\partial_t - \demi r \partial_r - \frac{x}{2} \nonumber \\
X_1 &=& - \frac{n}{2}t^2\partial_t\partial_r^{n-2} - t r \partial_r^{n-1} - \demi\mu r^2 - (x+\xi)t\partial_r^{n-2}
\nonumber \\
Y_{-\demi} &=& - \partial_r \label{2.1} \\
Y_{\demi} &=& - t\partial_r^{n-1} - \mu r \nonumber \\
M_0 &=& -\mu \nonumber
\EEA
and satisfy the commutators of $\mathfrak{age}(1)$, of which we give the non-vanishing ones
\BEQ \label{2.2}
\bigl[ X_0, Y_{\pm\demi}\bigr] \:=\: \mp\demi Y_{\pm\demi} \;\; , \;\; \bigl[ X_0, X_1 \bigr] \:=\: - X_1 \;\; , \;\;
\bigl[ Y_{\demi}, Y_{-\demi}\bigr] \:=\: M_0 .
\EEQ
However, there is a notable exception, namely
\BEQ
\bigl[ X_1, Y_{\demi}\bigr] \:=\: \frac{n-2}{2} t^2 \partial_r^{n-3} {\cal S}
\EEQ
with the Schr\"odinger operator
\BEQ
{\cal S} := n\mu \frac{\partial}{\partial t} - \frac{\partial^n}{\partial r^n}
+2\mu \left( x+\xi-\frac{n-1}{2}\right) \frac{1}{t}.
\EEQ
The generators (\ref{2.1}) form a dynamical symmetry of the Schr\"odinger equation ${\cal S}\psi(t,r)=0$, as can be seen
from the commutators
\BEQ
\bigl[ {\cal S}, Y_{\pm \demi}\big] \:=\: \bigl[ {\cal S}, M_0 \bigr] \:=\: 0 \;\; , \;\;
\bigl[ {\cal S}, X_0 \bigr] \:=\:  -\frac{n}{2}{\cal S} \;\; , \;\;
\bigl[{\cal S}, X_1\bigr] = -nt \partial_r^{n-2}{\cal S}.
\EEQ
In order to close the representation (\ref{2.1}), we must restrict the function space {\em modulo} solutions
of ${\cal S}\psi=0$.\footnote{This can be done by the following construction: to functions $f$ and $g$ are said to be
{\it equivalent}, written $f\sim g$, if there is a sufficiently differentiable function $\Lambda(t,r)$ such that
$f(t,r)=g(t,r)+\Lambda(t,r)$ and ${\cal S}\Lambda(t,r)=0$.} Then a natural function space for our purposes is
${\cal F} := \left.C^{1}\bigl( \mathbb{R}_+, C^n(\mathbb{R}) \bigr)\right/\sim$, the space of functions which are
continuously differentiable in time and $n$ times differentiable in space or alternatively
${\cal F} = \left.C^{1}\bigl( \mathbb{R}_+, H^n(\mathbb{R}) \bigr)\right/\sim$, where the Sobolev space $H^n(\mathbb{R})$
of $n$-times differentiable functions which together with their derivatives are also square-integrable is used
such that Fourier transforms with respect to
$r$ exist; and at the end with the quotient taken with
respect to the Schr\"odinger equation ${\cal S}\psi=0$.

Restricted to the space $\cal F$, the generators (\ref{2.1}) give for each integer $n> 2$ a non-local representation of
$\mathfrak{age}(1)$ which is a dynamical symmetry of the Schr\"odinger equation ${\cal S}\psi=0$.

\section{Finite transformations}

Besides the usual local generators of dilatations $X_0$, of spatial translations $Y_{-\demi}$ and of phase shifts $M_0$,
the representation (\ref{2.1}) also contains the
non-local generators $Y_{\demi}, X_{1}$ whose effect cannot be interpreted as a simple space-time coordinate transformation
$t\mapsto t'(t,r)$, $r\mapsto r'(t,r)$. On the other hand, we can still write
the formal Lie series $F(\eps,t,r) = e^{-\eps Y_{1/2}} F(0,t,r)$ and $F(\eps,t,r) = e^{-\eps X_{1}} F(0,t,r)$. They
are given as the solutions of the two initial-value problems
\BEA
\Bigl( \partial_{\eps} - t\partial_r^{n-1} -\mu r \Bigr) F(\eps,t,r) = 0 \;\; , \;\; F(0,t,r) = \phi(t,r)
\label{3.1} \\
\Bigl( \partial_{\eps} -\frac{n}{2} t^2\partial_t\partial_r^{n-2} -tr\partial_r^{n-1} -x t\partial_r^{n-2}
-\frac{1}{2}\mu r^2 \Bigr) F(\eps,t,r) = 0 \;\; , \;\; F(0,t,r) = \phi(t,r)
\label{3.2}
\EEA
such that the initial function $\phi\in{\cal F}$.

\begin{table}
\caption[Table 1]{Comparison of the finite transformations $e^{-\eps Y_{1/2}}\phi(t,r)$
for the generalised, non-local  Galilei-transformation when $z=n>2$ with the standard
local Galilei-transformation for $z=n=2$. The initial
distribution $\phi\in{\cal F}$ and $\mu=0$. \label{tab1}}
\begin{center}
\begin{tabular}{|l|ll|l|} \hline
$\phi(t,r)$ & non-local, $n>2$          & local, $n=2$                  & \\ \hline
$t^m$       & $t^m$                    & $t^m$                         & $m\in\mathbb{N}$ \\
$r^k$       & $r^k$                    & $(r+ t\eps)^k$                & $1\leq k \leq n-2$ \\
$r^{n-1}$   & $r^{n-1}+(n-1)!\, t\eps$ &  $\bigl( r+t\eps\bigr)^{n-1}$ & \\ \hline
\end{tabular}\end{center}
\end{table}
\begin{table}
\caption[Table 2]{Comparison of the finite transformations $e^{-\eps X_{1}}\phi(t,r)$
for the generalised, non-local special Schr\"odinger-transformation when $z=n=3$ or $4$
with the standard local special Schr\"odinger-transformation for $z=n=2$.
The initial distribution $\phi\in{\cal F}$ and $\mu=0$. \label{tab2}}
\begin{center}
\begin{tabular}{|l|ll|l|l|} \hline
            & \multicolumn{2}{|c|}{non-local}                                      & local              & \\
$\phi(t,r)$ & $n=3$                                    & $n=4$                     & $n=2$              & \\ \hline
$t^m$ & $t^m$                                          & $t^m$                     & $t^m/(1-t\eps)^{m+x+\xi}$ &
                                                                                     $m\in\mathbb{N}$\\
$r$   & $r+(x+\xi)t\epsilon$                           & $r$                       & $r/(1-t\eps)^{1+x+\xi}$ & \\
$r^2$ & $r^2+2(x+\xi+1)tr\eps$                         & $r^2 +2(x+\xi)t\eps$      & $r/(1-t\eps)^{2+x+\xi}$ & \\
      & ~~~$\:+\demi{(x+\xi+1)(2x+2\xi+3)}t^2\eps^2$   &                           &                         & \\
$r^3$ & ------                                         & $r^3 +6(x+\xi+1)tr\eps$\, & $r/(1-t\eps)^{3+x+\xi}$ & \\
\hline
\end{tabular}\end{center}
\end{table}

In tables~\ref{tab1} and ~\ref{tab2}, we illustrate these Lie series for the choices $\phi(t,r)=t^m$ and $\phi(t,r)=r^k$
with $m\in\mathbb{N}$ and $1\leq k\leq n-1$, which for $\mu=0$ solve the Schr\"odinger equation ${\cal S}\phi(t,r)=0$.
Comparison with the effects of the local Galilei- and special Schr\"odinger transformation shows important differences.
For example, although the spatial coordinate $r$ is left invariant by both generators when $n>2$, this does
not imply that these generators would not generate any spatial transformation, as we see from the transformation behaviour
of the higher powers of $r$. While in the local case $n=2$, the transformation of the powers $r^k$ is simply given
by taking the corresponding power of the transformation law of $r$ itself, this is no longer true in the non-local cases
$n>2$. While the action of the generators $Y_{\demi}$ and $X_1$, in our example, cannot be interpreted in terms of a
local coordinate transformation, the results look reminiscent to a transformation of a statistical distribution, where the
first moment happens to be invariant, but the higher ones change.
Therefore, these examples suggest that a better interpretation
might be to consider a transformation of an
initial distribution of spatial (or temporal) coordinates, where $\phi(t,r)$ would then take the r\^ole of a
distribution function.

In what follows, we give further results on the transformation of $\phi(t,r)$ and discuss possible consequences for
an interpretation. In order to keep the expressions to a manageable size, we shall concentrate on the two cases
$z=3$ and $z=4$. These are the values of $z$ in the Bray-Rutenberg theory of the growth of the relevant time-dependent
length scale $L(t)\sim t^{1/z}$ in O($n$)-symmetric systems with a conserved order parameter and quenched to $T<T_c$ 
\cite{Bray94a}. 

\subsection{The case $z=n=3$}

We now give the full transformation laws of the distribution $\phi(t,r)$.
We begin with the generalised Galilei transformation (\ref{3.1}) and use the
Fourier representation
\BEQ \label{3.3}
F(\eps,t,r) = \frac{1}{\sqrt{2\pi\,}\,} \int_{\mathbb{R}} \!\D k\: e^{\II k r} \wht{F}(\eps,t,k)
\EEQ
This leads to the equation $\bigl(\partial_{\eps}+t k^2 -\II \mu \partial_k \bigr) \wht{F}=0$. Letting $v :=\eps-\II k/\mu$
and $\wht{F}(\eps,t,k)=\wht{G}(v,t,k)$, it readily follows that
$\wht{G}(v,t,k) = \wht{G}_0(v,t)\exp\bigl(-\II t k^3/(3\mu)\bigr)$ and where $\wht{G}_0(v,t)$ must be found from the initial
condition, with the result $\wht{G}_0(v,t)  = \wht{\phi}(t,\II\mu k) \exp\bigl( t\mu^2 k^3/3\bigr)$. This gives the
transformed distribution in Fourier space
\BEQ \label{3.4}
\wht{\phi}(t,k) \mapsto \wht{F}(\eps,t,k) = \wht{\phi}(t,k+\II \mu\eps) \, \exp\left[-t k^2 \eps-\II \mu t k \eps^2 +
\frac{1}{3}\mu^2 t \eps^3 \right]
\EEQ
and finally in direct space, after having performed the integral over $k$, the general solution to eq.~(\ref{3.1}) becomes
\BEQ \label{3.5}
F(\eps,t,r) = \frac{1}{\sqrt{4\pi t \eps\,}\,} \int_{\mathbb{R}} \!\D r'\: \phi(t,r')
\exp\left[ -\frac{1}{4 t\eps} \left( \bigl( r-r'-t\mu\eps^2\bigr)^2 -4\mu t r' \eps^2
- \frac{4}{3} \mu^2 t^2 \eps^4 \right) \right].
\EEQ
Setting $\mu=0$, we obtain the entries in table~\ref{tab1}. Up to the $\mu$-dependent
terms, eq.~(\ref{3.5}) is a convolution of the initial distribution with a gaussian and using the form (\ref{3.4}), it is
readily checked that the group property holds true.

Specifically, we list some examples of finite transformations when $\mu\ne 0$:
\BEA
r'&=&(r+\mu t\epsilon^2)e^{\mu\epsilon r+{\mu t\epsilon^3}/{3}}\nonumber\\
(r^2)' &=&\left((r+\mu\epsilon^2t)^2+2t\epsilon\right) e^{\mu\epsilon r+{\mu t\epsilon^3}/{3}}\nonumber\\
\left(t+\frac{\mu}{2}r^3\right)' &=&
\left(t+\frac{\mu}{2}(r+\mu\epsilon^2t)^3+3\mu t\epsilon(r+\mu t\epsilon^2)\right) e^{\mu\epsilon r+{\mu t\epsilon^3}/{3}}.
\label{eq:3.6}
\EEA
We have checked that these solve (\ref{3.1}), as well as the Schr\"odinger equation with $n=3$.

In particular, if one tentatively
interprets $\phi(r)$ as a probability distribution such that $\int_{\mathbb{R}}\!\D r\: \phi(t,r)=1$,
this normalisation condition remains unchanged for $\mu=0$, viz.
$\left.\int_{\mathbb{R}}\!\D r\, F(\eps,t,r)\right|_{\mu=0}=1$. Furthermore, one may consider
\BEQ
\wht{\phi}(t,k) =  \left\langle e^{-\II k r}\right\rangle = \frac{1}{\sqrt{2\pi\,}\,}
\int_{\mathbb{R}} \!\D r\: e^{-\II k r} \phi(t,r)
\EEQ
as the associated characteristic function. For example, if we consider a shifted gaussian with characteristic function
$\wht{\phi}(t,k) = \exp\bigl( -\lambda k^2 +\II \gamma k\bigr)$, this transforms into
\BEQ
\wht{\phi}(t,k) \mapsto \wht{F}(\eps,t,k) = e^{-(\lambda+t\eps)k^2 +\II\gamma k} \,
e^{-\II k(2\mu\lambda \eps + \mu t \eps^2)} \,
e^{\mu^2 (\lambda+t\eps/3)\eps^2  -\mu \gamma \eps}
\EEQ
For $\mu=0$
the centre stays unchanged at $\gamma$, while the width becomes $\lambda \mapsto \lambda + t \eps$. Gaussian
distributions are therefore co-variant under the generalised Galilei generator $Y_{\demi}$ with $\mu=0$.
However, since the gaussian
distribution is not a solution of the Schr\"odinger equation with $n\ne 2$, one can realise a gaussian distribution at best
as an initial condition which has to be evolved in time. This illustrates the non-trivial constraint of remaining within
the reduced function space, introduced in section~2.

The integration of the generalised special transformation (\ref{3.2}) runs along similar lines.
For the sake of brevity, we set $\mu=0$ from now on.
In Fourier space, we introduce the new variables $u := \eps +2\II/(kt)$ and $v := k t^{2/3}$ and find
$\wht{F}(\eps,t,k) = \wht{G}(t,u,v) = t^{2(2-x-\xi)/3} \wht{G}_0(u,v)$. The as yet undetermined function
$\wht{G}_0$ is related to the initial distribution via
\BEQ
\wht{G}_0(\alpha,\beta) = \frac{1}{\sqrt{2\pi\,}\,} \left( \frac{2\II}{\alpha \beta}\right)^{2(x+\xi-2)}
\int_{\mathbb{R}} \!\D r'\: \exp\left( \frac{\II}{4} r' \alpha^2 \beta^3\right)
\phi\left( \left(\frac{2\II}{\alpha\beta}\right)^3, r'\right)
\EEQ
Hence, the final form for the solution of (\ref{3.2}) with $\mu=0$ reads
\BEQ \label{3.9}
F(\eps,t,r) = \frac{1}{2\pi} \int_{\mathbb{R}^2} \!\D k \D r'\: e^{\II k (r-r')}\,
\left( 1 + \frac{t k \eps}{2\II}\right)^{2(1-x-\xi)}
\phi\left( t \left(1+\frac{tk\eps}{2\II}\right)^{-3} , r' \left(1+\frac{tk\eps}{2\II}\right)^{-2} \right)
\EEQ
In particular, the entries in table~\ref{tab2} are recovered.

When we consider a gaussian distribution, we find the formal transformation
\BEQ
\wht{\phi}(t,k) = \sqrt{\frac{\lambda}{\pi}\,}\, e^{-\lambda k^2} \mapsto \wht{F}(\eps,t,k) =
\left( 1 +\frac{tk\eps}{2\II}\right)^{2(1-x-\xi)} \sqrt{\frac{\lambda_{\rm eff}(tk)}{\pi}\,}\,
e^{-\lambda_{\rm eff}(tk) k^2}
\EEQ
but now with a $k$-dependent effective width $\lambda_{\rm eff}(tk) = \lambda \bigl( 1 +tk\eps/(2\II)\bigr)^4$.
In contrast to the generalised Galilei transformation studied before, the transformation law also depends on the
value of the scaling dimension $x+\xi$ and we see that the companion factor reduces to unity, if $x+\xi=1$, that is
precisely when the time-dependent potential term in the Schr\"odinger equation ${\cal S}\psi=0$ vanishes. Again, a
gaussian distribution can at best be realised as an initial distribution.

Alternatively, one may implement the constraint of resting in the reduced function space of solutions of the
Schr\"odinger equation directly, which we now illustrate for $\mu\ne 0$. Together with eq.~(\ref{3.2}), we must require
the Schr\"odinger equation
\BEQ
{\cal S} F(\eps, t, r)=\left(3\mu\partial_t+(x+\xi-1)\frac{2\mu}{t}-\partial_r^3\right) F(\eps, t, r) \stackrel{!}{=}0.
\label{eq:auxiliar}
\EEQ
This system of equations is best solved in Fourier space, where we have
\BEA
\left( \partial_{\eps} -\frac{3\II}{2}k t^2 \partial_t + \II t k^2\partial_k - \II(x-2)tk +\frac{\mu}{2}\partial_k^2\right)
\wht{F}(\eps,t,k) &=& 0
\nonumber \\
\left( 3\mu\partial_t +\II k^3 + 2\mu (x+\xi-1) t^{-1} \right) \wht{F}(\eps,t,k) &=& 0
\label{3.13}
\EEA
together with the initial condition $\wht{F}(0,t,k) = \wht{\phi}(t,k)$.
The second of these is solved by
\BEQ \wht{F}(\eps, t, k)= t^{-2(x+\xi-1)/3}\: e^{-{\II k^3t}/{3\mu}}\: \wht{f}(\epsilon, k).
\label{eq:tdependence}
\EEQ
and the first condition (\ref{3.13}) then leads to a diffusion equation
\BEQ
\left(\partial_\epsilon+\frac{\mu}{2}\partial^2_k\right)\wht{f}(\epsilon, k)=0.
\label{eq:ffound}
\EEQ
where the diffusion constant is given by $-\mu/2$. Standard methods give the general solution $\wht{f}$ and using
(\ref{eq:tdependence}) we have formally
\BEQ
\wht{F}(\eps,t,k) = \frac{1}{\sqrt{-2\pi\mu\eps\,}} \int_{\mathbb{R}}\!\D \ell\:
\wht{F}(0,t,\ell) \exp\left[ \frac{(k-\ell)^2}{2\mu\eps} + \frac{\II t}{3\mu} (\ell^3 - k^3)\right]
\EEQ
Going back to direct space, we finally have (using analytic continuation where necessary)
\BEA
\lefteqn{
F(\eps,t,r) = \frac{1}{\sqrt{-8\pi^3 \mu\eps\,}} \int_{\mathbb{R}^3} \!\D r'\, \D k\, \D\ell\: \phi(t,r')
e^{\II(k-\ell)r +\II\ell(r-r')}
\exp\left[ \frac{(k-\ell)^2}{2\mu\eps} - \frac{\II t}{3\mu} (k-\ell)(k^2+k\ell + \ell^2)\right]
}
\nonumber \\
&=& \frac{1}{\sqrt{-8\pi^3 \mu\eps\,}} \int_{\mathbb{R}^2} \!\D r'\, \D m\: \phi(t,r'+r-tm^2/\mu)
e^{\II m (r-tm^2/3\mu) +m^2/(2\mu\eps)} \int_{\mathbb{R}}\!\D\ell\: \exp\left[ -\II \ell r' -\II \ell^2 t m/\mu\right]
\nonumber \\
&=& \frac{1}{\sqrt{-8\pi^2\II \eps t\,}} \int_{\mathbb{R}^2} \!\D r'\, \frac{\D m}{m^{1/2}}\; \phi(t,r'+r-tm^2/\mu)
\exp\left[ \II m \left(r - \frac{t}{3\mu} m^2\right) +\frac{\II\mu r'^2}{4 t m} +\frac{m^2}{2\mu\eps}\right]
\label{3.17}
\EEA
where in the last step the Fresnel integrals were used.
We remark that the scaling dimension $x+\xi$ does not appear explicitly.

For illustration, we write down the transformed time, which can be derived as follows. From (\ref{3.17}),
we have with $\phi(t,r)=t$, carrying out
first the integral over $r'$ via a Fresnel integral
\BEA
t' &=& \frac{t}{\sqrt{-8\pi^2\II\eps t\,}} \int_{\mathbb{R}} \!\D r'\: e^{\II \mu/(4 tm) r'^2}
\int_{\mathbb{R}} \!\frac{\D m}{m^{1/2}}\; e^{\II m(r -(t/3\mu)m^2) +m^2/(2\mu\eps)}
\nonumber \\
&=& \frac{t}{\sqrt{2\pi\mu\eps\,}}
\int_{\mathbb{R}} \!\D m\: \exp\left[ -\frac{\II t}{3\mu}
\left( m^3 - \frac{3}{2\II} \frac{1}{t\eps} m^2 - \frac{3\mu r}{t} m\right)\right]
\nonumber \\
&=& t \; \sqrt{\frac{2\pi}{\mu \eps}\,}
\left(\frac{\mu}{t}\right)^{1/3}
\exp\left[ \frac{1}{2t\eps} \left(r - \frac{1}{6\mu t\eps^2}\right)\right]
\mbox{\rm Ai}\left( - \left(\frac{\mu}{t}\right)^{1/3} \left(r - \frac{1}{4\mu t \eps^2}\right)\right)
\EEA
where $\mbox{\rm Ai}$ is the Airy function and in the second line we performed a shift in the integration variable
in order to eliminate the terms $\sim m^2$ in the exponential.\\

\subsection{The case $z=n=4$}

The finite form of the generalised Galilei transformation is found by solving (\ref{3.1}) along the same lines
as for the case $z=3$. In Fourier space, we obtain
\BEQ
\wht{\phi}(t,k) \mapsto {\wht F}(\eps, t, k) = e^{t\mu^3\eps^4/4}\exp\left(\II
t\mu^2\eps^3k+\frac{3}{2}t\mu\eps^2 k^2-\II t\eps k^3\right){\wht{\phi}}(t, k+\II\mu\eps)
\EEQ
quite analogous to (\ref{3.4}). From this, we find in direct space
\BEA
F(\eps,t,r)=
\frac{\exp({t\mu^3\eps^4/4})}{2\pi}\int_{\mathbb{R}}\!\D r'{\;}\phi(t,r') \, e^{\mu\eps r'}
\int_{\mathbb{R}} \!\D k{\;}
e^{\II k (r-r'+t\mu^2\eps^3)+\frac{3}{2}t\mu\eps^2 k^2-\II t\eps k^3}.\label{eq:4galfinresult}
\EEA
Setting $\mu=0$, the results in table~\ref{tab1} can be recovered and we also have the same conservation of the
normalisation, when $\mu=0$. Some explicit examples for transformations with $\mu\ne 0$ read
\BEA
r'                &=& (r+\mu^2 t\epsilon^3)e^{\mu\epsilon r+{\mu^3 t\epsilon^4}/{4}}
\nonumber\\
\left(r^2\right)' &=& \left((r+\mu^2\epsilon^3t)^2+3\mu t\epsilon^2\right) e^{\mu\epsilon r+{\mu^3 t\epsilon^4}/{4}}
\nonumber \\
\left(r^3\right)' &=& \left((r+\mu^2\epsilon^3t)^3+9\mu t\epsilon^2(r+\mu^2 t\epsilon^3)+6t\epsilon\right)
e^{\mu\epsilon r+{\mu^3 t\epsilon^4}/{4}}.
\label{eq:z4galtransfcord}
\EEA

Next, we integrate the special generator $X_1$ by solving (\ref{3.2}). Again, this is best done in Fourier space and
we set $\mu=0$ for brevity. Since with
respect to the case $n=3$, some subtleties arise, we proceed step by step. First, we introduce the new variable
$v=k t^{1/2}$ and set $\wht{F}(\eps,t,k) = \wht{G}(\eps,t,v)$ which satisfies the equation
\BEQ
\bigl(\partial_{\eps}+2v^2t\partial_t+(x+\xi-3)v^2\bigr)\wht{G}(\eps, t, v)=0
\EEQ
This in turn is solved by setting $u=\eps-{\ln t}/(2 y^2)$ and we find
\BEQ
\wht{G}(\eps, t, v)= \wht{g}_0(u,v)t^{-(x+\xi-3)/2}
\EEQ
where $\wht{g}_0(u,v)$ is determined from the initial condition
\BEQ
{\hat g}_0(u, v)= e^{-(x+\xi-3)uv^2}\wht{\phi}\left(e^{-2uv^2}, e^{uv^2}v\right).
\EEQ
Using the inverse transformations $t=e^{-2uv^2}$, $k=e^{uv^2}v$, the final form is
\BEQ
\wht{\phi}(t,k) \mapsto \wht{F}(\eps, t, k)= e^{-(x+\xi-3)\eps tk^2}\:
\wht{\phi}\left(e^{-2\eps tk^2}t, e^{\eps tk^2}k \right).
\EEQ
and this gives in direct space (with $\mu=0$)
\BEQ
F(\eps,t,r) = \frac{1}{2\pi}\int_{\mathbb{R}^2}\!\D k\D r'{\;}
e^{\II k(r-r')-(x+\xi-2)\eps tk^2} \: \phi\left(e^{-2\eps tk^2}t, e^{-\eps tk^2}r'\right)
\EEQ
from which the corresponding entries in table~\ref{tab2} follow.\footnote{All entries in tables~\ref{tab1} and~\ref{tab2}
can be checked by direct substitution.} The main difference with respect to (\ref{3.9}) is the
exponential rescaling of time and space. The case $\mu\ne 0$ can be treated in the same way as $n=3$ case.
We omit the calculation.

\section{Covariant two-point functions}

We now derive the form of the co-variant two-point function
\BEQ
F = F(t_1,t_2;r_1,r_2) = \left\langle \phi_1(t_1,r_1) \phi_2(t_2,r_2) \right\rangle
\EEQ
and where the scaling operators $\phi_i$ have scaling dimension $(x_i,\xi_i)$ and mass $\mu_i$.
The co-variance of $F$ is expressed by the conditions $X^{(2)} F=0$, where $X^{(2)}$ is the two-body
extension of the generators $X\in\mathfrak{age}(1)$ constructed in section~2.

Spatial translation-invariance $Y_{-\demi} F=0$ leads to $F=F(t_1,t_2,r)$, with $r=r_1-r_2$.
The mass-invariance $M_0 F=0$ gives the Bargman super-selection rule $\mu_1+\mu_2=0$. The requirement of generalised
Galilei-invariance $Y_{\demi}F\stackrel{!}{=}0$ leads to, using again the Bargman super-selection rule
\BEQ
Y_{\demi} F = \left( -t_1 \frac{\partial^{n-1}}{\partial r_1^{n-1}} - \mu_1 r_1
-t_2 \frac{\partial^{n-1}}{\partial r_2^{n-1}} - \mu_2 r_2 \right) F =
\left( - \left(t_1 + (-1)^{n-1} t_2\right) \frac{\partial^{n-1}}{\partial r^{n-1}} - \mu_1 r \right) F = 0
\EEQ
It follows that one must distinguish between the cases (i) $n$ even and (ii) $n$ odd. \\

\noindent
\underline{\bf 1. $n$ even}. We rewrite the two-point function as
\BEQ
F = F(u,v,r) \;\; , \;\; u := t_1 - t_2 \;\; , \;\; v := t_1 / t_2 \;\; , \;\; r := r_1 - r_2
\EEQ
and obtain from the three co-variance conditions $Y_{\demi}F=0$, $X_0F=0$ and $X_1 F=0$ the equations
\BEA
\left[ -  u\partial_r^{n-1} - \mu_1 r \right] F &=& 0 \label{4.4} \\
\left[ - \frac{n}{2} u \partial_u - \frac{1}{2} r \partial_r - \frac{x_1 + x_2}{2} \right] F &=& 0 \label{4.5} \\
\left[ - \frac{n}{2} u^2 \frac{v+1}{v-1} \partial_u \partial_r^{n-2} - \frac{n}{2} u v \partial_v \partial_r^{n-2}
- \frac{uv}{v-1} r \partial_r^{n-1}  \right. & & \nonumber \\
\left. - (x_1+\xi_1)\frac{uv}{v-1}\partial_r^{n-2} - (x_2+\xi_2)\frac{u}{v-1}\partial_r^{n-2}
-\demi\mu_1 r^2 \right] F &=& 0 \label{4.6}
\EEA
Acting with $\partial_r^{n-2}$ on (\ref{4.5}), eq.~(\ref{4.6}) can be simplified to
\BEQ
\partial_r^{n-2} \left( n v \partial_v + \frac{v}{v-1}\left( x_1 - x_2 +2\xi_1 -n+2\right) +
\frac{1}{v-1}\left( x_2 - x_1 +2\xi_2 -n+2\right) \right) F = 0 \label{4.7}
\EEQ
It clear that each of the equations (\ref{4.4},\ref{4.5},\ref{4.7}) will fix the dependence of $F=F(u,v,r)$ on one of
its variables. In fact, the scaling form obtained from eqs.~(\ref{4.4},\ref{4.5}) implies that the dependence of $F$ on $v$
factorises such that the $(n-2)$-fold derivative in
(\ref{4.7}) can be dropped. It can also be explicitly checked that the closure condition of our representation is
automatically satisfied, as it should be. We find the following scaling form
\BEQ \label{4.8}
F(u,v,r) = t_2^{-(x_1+x_2)/n}\: (v-1)^{-\frac{2}{n}[ (x_1+x_2)/2+\xi_1+\xi_2-n+2]} \:
v^{-\frac{1}{n}[ x_2-x_1+2\xi_2-n+2]}\: f\left( r u^{-1/n}\right)
\EEQ
where the form of the last scaling function $f=f(y)$ follows form eq.~(\ref{4.4})
\BEQ \label{4.9}
\frac{\D^{n-1} f(y)}{\D y^{n-1}} + \mu_1 y f(y) = 0
\EEQ

\noindent
\underline{\bf 2. $n$ odd}. We rewrite the two-point function as
\BEQ
F = F(u,v,r) \;\; , \;\; u := t_1 + t_2 \;\; , \;\; v := t_1 / t_2 \;\; , \;\; r := r_1 - r_2
\EEQ
and now obtain from the three co-variance conditions $Y_{\demi}F=0$, $X_0F=0$ and $X_1 F=0$ again the equations
(\ref{4.4},\ref{4.5}), of course with the modified relationship between $u$ and $t_{1,2}$, while (\ref{4.6}) is replaced by
\BEA
\left[ - \frac{n}{2} u^2 \frac{v-1}{v+1} \partial_u \partial_r^{n-2} - \frac{n}{2} u v \partial_v \partial_r^{n-2}
- \frac{uv}{v+1} r \partial_r^{n-1}  \right. & & \nonumber \\
\left. - (x_1+\xi_1)\frac{uv}{v+1}\partial_r^{n-2} - (x_2+\xi_2)\frac{u}{v+1}\partial_r^{n-2}
-\demi\mu_1 r^2 \right] F &=& 0 \label{4.11}
\EEA
Using again (\ref{4.5}), we find the more simple condition
\BEQ
\partial_r^{n-2} \left( n v \partial_v + \frac{v}{v+1}\left( x_1 - x_2 +2\xi_1 -n+2\right) +
\frac{1}{v+1}\left( x_2 - x_1 +2\xi_2 -n+2\right) \right) F = 0 \label{4.12}
\EEQ
This leads to the scaling form
\BEQ \label{4.13}
F(u,v,r) = t_2^{-(x_1+x_2)/n} \:(v+1)^{-\frac{2}{n}[ (x_1+x_2)/2+\xi_1+\xi_2-n+2]} \:
v^{-\frac{2}{n}[ x_2-x_1+\xi_1-\xi_2]} \:f\left( r u^{-1/n}\right)
\EEQ
and where the scaling function $f(y)$ is again given by eq.~(\ref{4.9}).

It remains to discuss the remaining scaling function $f(y)$. The general solution of (\ref{4.9}) is
\BEQ
f(y) = \sum_{\ell=0}^{n-2} f_{\ell}\, y^{\ell}\, {}_1F_{n-1}\left( 1;
\frac{2+\ell}{n},\frac{3+\ell}{n},\ldots,\frac{n+\ell}{n}; -\frac{ \mu_1 y^n}{n^{n-1}} \right)
\EEQ
where $_1F_{n-1}$ are generalised hyper-geometric functions and the $f_{\ell}$ are normalisation constants. On this,
physically reasonable boundary conditions must be imposed, especially $\lim_{y\to\infty} f(y)=0$.
It may be more instructive, however, to look at explicit examples.

\begin{figure}[tb]
\centerline{\psfig{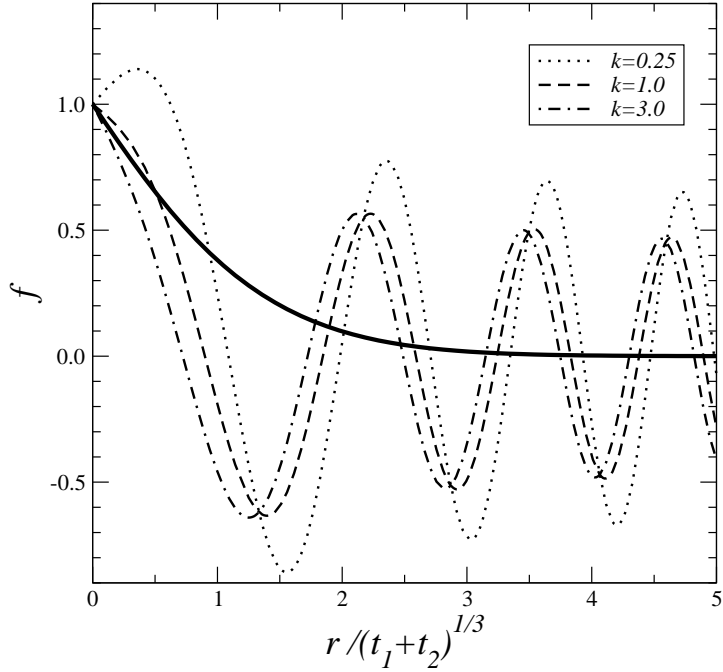}}
\caption{\label{fig:n=3} Scaling function $f(y)$ in the case $z=n=3$, normalised to $f(0)=1$.
The solid line gives the behaviour if
$\mu_1=1>0$, while the broken lines indicate the behaviour, for $\mu_1=-8<0$ and several values of $k$,
of the function
$f(y)=(\mbox{\rm Ai}(|\mu_1|^{1/3} y) +k \mbox{\rm Bi}(|\mu_1|^{1/3} y))/(\mbox{\rm Ai}(0)) +k \mbox{\rm Bi}(0))$.}
\end{figure}

\noindent
\underline{\bf 1. $n=3$}. In this case, eq.~(\ref{4.9}) reduces essentially to Airy's equation and the solutions can be
compactly expressed in terms of Airy's functions and the normalisation constants $f_{1,2}$
\BEA
f(y) = f_1 \mbox{\rm Ai}\left( - \mu_1^{1/3} y\right) &;& \mu_1 >0 \nonumber \\
f(y) = f_1 \mbox{\rm Ai}\left(  |\mu_1|^{1/3} y\right) +  f_2 \mbox{\rm Bi}\left( |\mu_1|^{1/3} y\right) &;& \mu_1 <0
\label{4.15}
\EEA
For $\mu_1>0$, the  second independent solution of (\ref{4.9}) was suppressed, since it diverges for $y\to\infty$.
Figure~\ref{fig:n=3} illustrates the behaviour of the scaling function, for positive and negative values of $\mu_1$. 
We discuss the shape of the scaling functions below. 

\begin{figure}[tb]
\centerline{\psfig{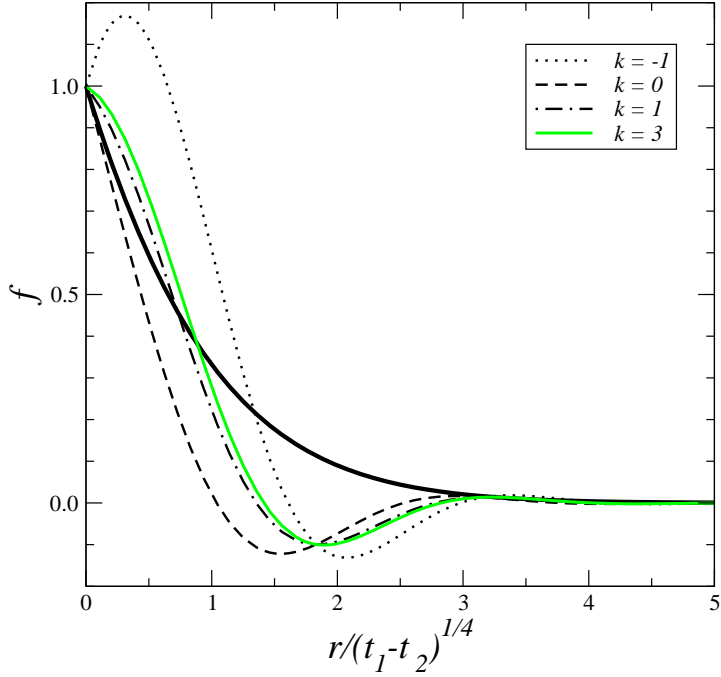}}
\caption{\label{fig:n=4} Scaling function $f(y)$ in the case $z=n=4$, normalised to $f(0)=1$.
The thick solid line gives the behaviour if $\mu_1=1>0$.  The broken lines and the grey line indicate the behaviour,
for $\mu_1=-8<0$ and several values of $k$, of the function
$f(y)=({\cal F}_1(y) + k {\cal F}_2(y))/({\cal F}_1(0) + k {\cal F}_2(0))$, with the ${\cal F}_i(y)$ defined in
(\ref{4.19}). }
\end{figure}

\noindent
\underline{\bf 2. $n=4$}. The solution of (\ref{4.9}) now takes the more simple form
\BEQ
f(y) = f_0\:{}_0F_2\left(\frac{1}{2},\frac{3}{4}; -\frac{\mu_1 y^4}{64}\right)
+ f_1\,y\: {}_0F_2\left(\frac{3}{4},\frac{5}{4}; -\frac{\mu_1 y^4}{64}\right)
+ f_2\,y^2\: {}_0F_2\left(\frac{5}{4},\frac{3}{2}; -\frac{\mu_1 y^4}{64}\right)
\EEQ
This may be analysed using the leading asymptotic behaviour of the hyper-geometric function $_0F_2$, which may be read off
from Wright's formul{\ae} \cite{Wright35}
\BEA
{}_0F_2(a,b;z) &\stackrel{z\to\infty}{\simeq}& \frac{\Gamma(a)\Gamma(b)}{2\pi\sqrt{3\,}}\: z^{(1-a-b)/3}\; e^{3 z^{1/3}}
\nonumber \\
{}_0F_2(a,b;-z) &\stackrel{z\to\infty}{\simeq}& \frac{\Gamma(a)\Gamma(b)}{\pi\sqrt{3\,}}\: z^{(1-a-b)/3}\; e^{3 z^{1/3}/2}
\cos\left( \frac{3\sqrt{3\,}}{2} z^{1/3} +\frac{\pi}{3}(1-a-b)\right)
\label{4.17}
\EEA
For both $\mu_1>0$ and $\mu_1<0$, this implies that the function $f(y)$ diverges exponentially fast as $y\to\infty$.
We absorb this divergence by choosing the constants $f_{0,1,2}$ accordingly and then find
\BEA
f(y) = f_0 \left[ {}_0F_2\left(\frac{1}{2},\frac{3}{4};-\frac{\mu_1 y^4}{64}\right)
-\frac{\sqrt{2\,}\,\Gamma(3/4)}{\Gamma(1/2)} \mu_1^{1/4} y\:
{}_0F_2\left(\frac{3}{4},\frac{5}{4};-\frac{\mu_1 y^4}{64}\right)
\right.
& & \nonumber \\
\left. +\frac{\Gamma(3/4)}{\Gamma(1/4)} \mu_1^{1/2} y^2\:
{}_0F_2\left(\frac{5}{4},\frac{3}{2};-\frac{\mu_1 y^4}{64}\right) \right] &;& \mu_1 >0
\label{4.18} \\
f(y) = f_{(0)} \bigl[ {\cal F}_1(y) + k {\cal F}_2(y) \bigr] \hspace{7.6truecm} &;& \mu_1 <0
\nonumber
\EEA
where $f_0$ and $f_{(0)}$ are normalisation constants, $k$ is a free parameter and
\BEA
{\cal F}_1(y) &:=& |\mu_1|^{1/4} y\: {}_0F_2\left(\frac{3}{4},\frac{5}{4};\frac{|\mu_1| y^4}{64}\right)
- \frac{\Gamma(1/2}{\Gamma(3/4)}\: {}_0F_2\left(\frac{1}{2},\frac{3}{4};\frac{|\mu_1| y^4}{64}\right)
\nonumber \\
{\cal F}_2(y) &:=& |\mu_1|^{1/2} y^2\: {}_0F_2\left(\frac{5}{4},\frac{3}{2};\frac{|\mu_1| y^4}{64}\right)
- \frac{\Gamma(1/4}{\Gamma(3/4)}\: {}_0F_2\left(\frac{1}{2},\frac{3}{4};\frac{|\mu_1| y^4}{64}\right)
\label{4.19}
\EEA
The behaviour of these scaling functions is illustrated in figure~\ref{fig:n=4}. We observe that once
having eliminated the asymptotically leading term eq.~(\ref{4.17}), the physically required boundary condition
$\lim_{y\to\infty} f(y) = 0$ is satisfied.

Comparing figures~\ref{fig:n=3} and~\ref{fig:n=4}, we notice that although the scaling function
satisfies for both $n$ odd and $n$ even the same differential equation (\ref{4.9}), the interpretation of the
scaling variable $|\mu_1|^{1/4} y$ is different. Indeed, for $n=3$, the time difference $t_1-t_2$ enters into $y$, whereas
for $n=4$ it is the sum $t_1+t_2$. 
Furthermore, we see that for $\mu_1>0$, only a single independent solution
remains, which decreases from $f(0)=1$ monotonously and very rapidly towards zero when $y$ is increased.
On the other hand, for $\mu_1<0$, we find two independent admissible solutions whose decay towards zero
is an oscillatory function of $y$ (for $n=3$, the decay should be algebraic, whereas it looks to be
(stretched) exponential for $n=4$). This feature may allow to distinguish at least qualitatively between two
physically distinct situations with $z>2$:
\begin{itemize}
\item non-equilibrium relaxation kinetics with a {\em conserved} order-parameter (model B dynamics).
Below the critical point, viz. $T<T_c$,
in systems with a global O($n$)-symmetry it is known that $z=3$ for a scalar order-parameter ($n=1$),
and $z=4$ for vector order-parameters ($n\geq 2$) \cite{Bray94a}. At criticality 
$z=4-\eta=4-\demi\frac{n+2}{(n+8)^2}\eps^2 +{\rm O}(\eps^3)$ in $d=4-\eps$ dimensions \cite{Zinn02}.
In these cases, scaling functions are generically seen to be oscillating.\footnote{For example, the scaling function
${\cal F}_2(y)$ in (\ref{4.19}) reproduces the exactly known two-time response in the $3D$ Mullins-Herring model of
surface growth with a conserved order-parameter \cite{Roethlein06}.}
\item in critical dynamics, viz. $T=T_c$, and without any conservation law on the order-parameter (model A dynamics),
the dynamical exponent $z\gtrsim 2$ \cite{Zinn02}. Here, the decay of the scaling functions
is in general monotonous.
\end{itemize}
Our results suggest that these physically distinct cases, even with the same value of $z$, 
might be distinguished through the {\em sign} of the
dimensionful parameter $\mu_1$, such that $\mu_1>0$ reproduces the monotonous decay seen in critical dynamics (model A)
whereas $\mu_1<0$ leads to the oscillatory decay found in conserved systems (model B).

\section{Conclusions}

This work has been motivated by the persistent difficulties to construct non-trivial Lie algebras of
space-time transformations. We believe that the possibility of finding Lie algebra generators which cannot be
expressed as vector fields merits serious consideration. We have constructed 
new representations of the ageing algebra $\mathfrak{age}(1)$, corresponding
to an integer dynamical exponent $z=n\geq 2$ to explore the mathematical structure of dynamical symmetries
whose infinitesimal generators are no longer described by the usual vector fields involving only first-order
differential operators. Provided that we restrict the admissible function space to the solution space of the
Schr\"odinger equation ${\cal S}\psi=0$, and thereby somewhat relax the requirements of a dynamical symmetry,
we have given an explicit $n$-dependent family of linear partial differential equations which are indeed
$\mathfrak{age}(1)$-invariant in the sense introduced here. An important open question is how to extend this to non-linear 
equations. 

The non-local infinitesimal generators of $\mathfrak{age}(1)$ contain higher-order differential operators. 
Their exponentiation does not lead to local spatio-temporal coordinate transformations and we have
considered the possibility that a better interpretation might be formulated in terms of transformation rules
for distributions of spatio-temporal coordinates. Several examples of such transformation rules have been derived.

Finally, we also studied the scaling form of co-variant two-point functions. Surprisingly, for $z=n$ {\em even}
the scaling forms are compatible with the expectations of a two-time {\em response} function (as it is usually
the case in present theories of local scale-invariance in ageing systems) since they depend on the time difference 
$t_1-t_2$. On the other hand, this is not so for $z=n$ {\em odd}, where
the arguments of the scaling functions are much more reminiscent of co-variant two-time {\em correlators}, since they
contain the sum $t_1+t_2$.
We have also seen that the shape of the space-dependent part of the scaling functions can at least qualitatively
account for the different forms found for non-conserved (model A) dynamics, where one expects a monotonous decay,
and for conserved (model B) dynamics, where scaling functions are oscillatory.. This is achieved through a simple change in 
the sign of the dimensionful `mass parameter' $\mu$. Although we think it unlikely that our non-local representations of
$\mathfrak{age}(1)$ should be directly applicable to physical models, we consider this qualitative feature
encouraging.

\appsektion{On Galilei-transformations}

For comparison with the non-local generators treated in the main text, we recall
the computation of finite Galilei-transformations, for distributions with a non-vanishing mass
$\mu\ne 0$. The infinitesimal generator is $Y_{1/2}=-t\partial_r -\mu r$, from which
the finite transformation is formally obtained as $F(\eps,t,r)=e^{-\eps Y_{1/2}} F(0,t,r)$.
It is given by the differential equation
\BEQ \label{a1}
\bigl( \partial_{\eps} - t\partial_r - \mu r\bigr) F(\eps,t,r) = 0 \;\; , \;\;
F(0,t,r) = \phi(t,r)
\EEQ
where $\phi=\phi(t,r)$ denotes the given initial distribution. Eq.~(\ref{a1})
is solved in Fourier space:
\BEQ
\wht{\phi}(t,k) \mapsto \wht{F}(\eps,t,k) =
\wht{\phi}(t,k+\II\mu\eps) \exp\left[ -\frac{1}{2}\mu t\eps + \II tk\eps\right]
\EEQ
In direct space, the galilei-transformed distribution becomes
\BEA \phi(t,r) \mapsto F(\eps,t,r) &=&
\frac{1}{\sqrt{2\pi\,}\,} \int_{\mathbb{R}} \!\D k\: \wht{\phi}(t,k+\II\mu\eps)\,
e^{\II k(r+t\eps)} \, e^{\mu t \eps^2 /2}
\nonumber \\
&=& \frac{1}{2\pi} \int_{\mathbb{R}} \!\D r'\: \phi(t,r) \, e^{\mu r'\eps - \mu t \eps^2 /2}
\underbrace{\int_{\mathbb{R}}\!\D k\: e^{\II k( r-r'+t \eps)}}_{2\pi\delta(r+t \eps -r')}
\nonumber \\
&=& \phi(t,r+ t \eps) \, e^{\mu(r+t\eps)\eps -\mu t\eps^2/2}
\EEA
Hence, for $\mu=0$, the initial distribution is rigidly shifted according to $t\mapsto t$ and
$r\mapsto r+ t\eps$. This is a consequence of
the local nature of the standard Galilei-transformation, which can be expressed in terms of a vector field.
In particular, the entries in table~\ref{tab1} are recovered.

\noindent
{\bf Acknowledgements:} Most of the work on this paper was done during the visits of S.S. at
the Universit\'e Henri Poincar\'e Nancy I.
S.S. is supported in part by the Bulgarian NSF grant {\it DO 02-257}.


\end{document}